\documentclass[conference]{IEEEtran}
\IEEEoverridecommandlockouts
\usepackage{cite}
\usepackage{amsmath,amssymb,amsfonts}
\usepackage{algorithmic}
\usepackage{graphicx}
\usepackage{textcomp}
\usepackage{xcolor}

\usepackage{algorithm}
\usepackage{algorithmic}
\usepackage{verbatim}
\def\BibTeX{{\rm B\kern-.05em{\sc i\kern-.025em b}\kern-.08em
    T\kern-.1667em\lower.7ex\hbox{E}\kern-.125emX}}
\begin{document}

\title{Computing Resource Allocation of Mobile Edge Computing Networks Based on Potential Game Theory}

%\author{\IEEEauthorblockN{1\textsuperscript{st} Heng Liu}
%\IEEEauthorblockA{\textit{School of Electronic Information and Communications} \\
%\textit{Huazhong University of Science and Technology}\\
%Wuhan, China \\
%h\_liu@hust.edu.cn}
%\and
%\IEEEauthorblockN{2\textsuperscript{nd} Haoming Jia}
%\IEEEauthorblockA{\textit{School of Electronic Information and Communications} \\
%\textit{Huazhong University of Science and Technology}\\
%Wuhan, China \\
%M201571777@hust.edu.cn}
%\and
%\IEEEauthorblockN{3\textsuperscript{rd} Jiaqi Chen}
%\IEEEauthorblockA{\textit{School of Electronic Information and Communications} \\
%\textit{Huazhong University of Science and Technology}\\
%Wuhan, China \\
%chenjq\_hust@hust.edu.cn}
%\and
%\IEEEauthorblockN{4\textsuperscript{th} Wanyue Xu}
%\IEEEauthorblockA{\textit{School of Electronic Information and Communications} \\
%\textit{Huazhong University of Science and Technology}\\
%Wuhan, China \\
%M201871948@hust.edu.cn}
%
%}

%\author{
%Heng Liu,
%Haoming Jia,
%Jiaqi Chen,
%Wanyue Xu \\
%
%School of Electronic Information and Communications \\
%Huazhong University of Science and Technology\\
%Wuhan, China \\
%E-mail:\{h\_liu, M201571777, chenjq\_hust, M201871948\}@hust.edu.cn
%}
\author{\IEEEauthorblockN{Heng Liu\IEEEauthorrefmark{2},
Haoming Jia\IEEEauthorrefmark{2},
Jiaqi Chen\IEEEauthorrefmark{2},
Xiaohu Ge\IEEEauthorrefmark{2}\IEEEauthorrefmark{1},
Yonghui Li\IEEEauthorrefmark{3},
Lin Tian\IEEEauthorrefmark{4},
Jinglin Shi\IEEEauthorrefmark{4}\IEEEauthorrefmark{5}
}
\IEEEauthorblockA{\IEEEauthorrefmark{2}School of Electronic Information and Communications\\ Huazhong University of Science and Technology, Wuhan, China}
\IEEEauthorblockA{\IEEEauthorrefmark{3}School of Electrical and Information Engineering\\
University of Sydney, Sydney, NSW 2006, Australia}
\IEEEauthorblockA{\IEEEauthorrefmark{4}Beijing Key Laboratory of Mobile Computing and Pervasive Devices\\ Institute of Computing Technology, Chinese Academy of Sciences, China}
\IEEEauthorblockA{\IEEEauthorrefmark{5}University of Chinese Academy of Sciences, China}
\IEEEauthorblockA{Contact Email: xhge@mail.hust.edu.cn}}

\maketitle

\begin{abstract}
 Mobile edge computing (MEC) networks are one of the key technologies for ultra-reliability and low-latency communications. The computing resource allocation solution needs to be carefully designed  to guarantee the computing resource efficiency of MEC networks. Based on the potential game theory, a computing resource allocation solution is proposed to reduce energy consumption and improve computing resource efficiency in MEC networks. The computing resource allocation solution includes two parts: the first part is the power control scheme based on the potential game theory and the second part is the computing resource allocation scheme based on linear programming. The power control scheme is to find a set of the transmission powers of base stations (BSs) that maximizes the potential function of MEC networks. The computing resource allocation scheme is to maximize the average computing resource allocation coefficient of the MEC networks based on the results of the power control scheme. Compared with traditional solutions, simulation results indicate the computing resource utilization and energy efficiency of the proposed computing resource allocation solution are significantly improved.
\end{abstract}


\section{Introduction}
The exponential increment of data traffic, the sustained growth of terminals, as well as more and more diverse service scenarios, increase the pressure of the fourth generation (4G) cellular networks, which has led to the advent of the fifth generation (5G) cellular networks \cite{1Ge}\cite{2Ge}. The MEC technology is a promising solution for 5G networks, which can provide the complex computing capability at the radio access network (RAN) \cite{3Yu}\cite{4Hu}. The function of the cloud data center is sunk to the edge of cellular networks by MEC technologies, which provide users with some functions of the core network such as computing, storage and communication resources in base stations (BSs) at the edge of wireless networks. However, the interference among adjacent BSs not only influences wireless traffic transmissions but also affects the resource allocation in MEC networks. It is an important challenge for resource allocation optimization in MEC networks.

In the literature about resource allocation in MEC networks, a joint caching and offloading mechanism was proposed to upload uncached computation results as well as download computation result at BSs \cite{5Cui}. However, this mechanism is not suitable for applications involving with large computational demands and time-critical requirements as well as large-scale computational results, such as augmented reality, interactive online gaming, and multimedia conversions. An energy-efficient resource allocation scheme was presented for a multi-user MEC system with inelastic computation tasks and non-negligible task execution durations \cite{6Guo}. However, this scheme only focuses on reducing energy consumption and neglects resource allocation coefficient in MEC systems. A fair resource-allocation scheme was proposed to maximize the total throughput of a wireless network when each users' transmission rate is constrained with the minimum transmission rate \cite{7Zhu}. Based on the double-sided auction game, an efficient resource allocation scheme with limited resources between suppliers and consumers was proposed in \cite{8Zou}. A new distributed resources block (RB) and power allocation (PA) algorithm based on non-cooperative game theory was presented to improve the energy efficiency of MEC networks \cite{9Sam}. However, the computing resource allocation utilization and energy efficiency of MEC networks have not been simultaneously investigated in existing studies.

This paper focuses on reducing the energy consumption and improving the computing resource allocation utilization in MEC networks. The main contributions of this paper is summarized as follows:
\begin{enumerate}
  \item To improve the computing resources utilization and energy efficiency of MEC networks, a computing resource allocation solution are proposed in this paper. Simulation results show that the proposed solution can save energy consumption and improve resource utilizations.
  \item To reduces the energy consumption of MEC networks, a power control scheme is developed based on the potential game theory.
  \item To improve the average computing resource allocation coefficient of MEC networks, a new computing resource allocation scheme is developed based on the linear programming.
\end{enumerate}

The rest of this paper is arranged as follows. Section II gives the system model. Section III describes the proposed computing resource allocation solution. Simulation results and analysis is presented in Section IV. Section V draws the conclusions.

\section{SYSTEM MODEL}

\begin{figure}[h]
\centerline{\includegraphics[width=0.35\textwidth]{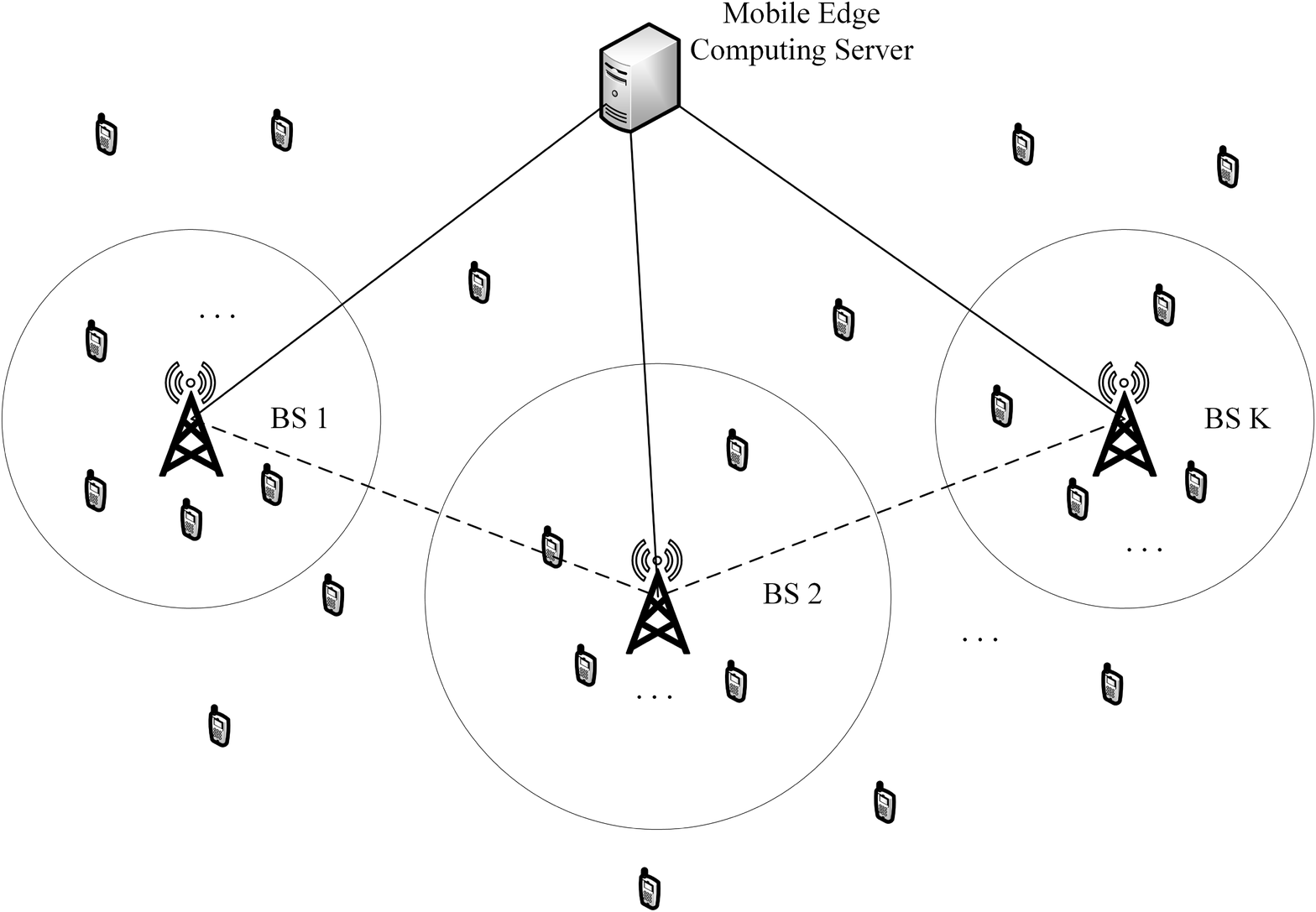}}
\caption{Mobile edge computing system model.}
\label{fig1}
\end{figure}

Without loss of generality, one MEC server and $K$ BSs are configured for a MEC network. $P_k$ denotes the transmission power of  the $k-\text{th}$ BS ( $k\in B=\{1,2,...,K\}$ ). The maximum transmission power of each BS is denoted as $P_{max}$. All users are assumed to be governed by a uniformly distribution with the density $\rho$. The coverage of the $k-\text{th}$ BS is configured as a circle with the radius of $r_k$. The maximum coverage radius is denoted as $r_M$ when the transmission power is $P_{max}$. The distance between the $m-\text{th}$ BS and the $n-\text{th}$ BS is denoted by $R_{mn}$. The required computing resources for the $k-\text{th}$ BS is denoted by $f^{BS}_k$ and the computing resources actually allocated to the $k-\text{th}$ BS is denoted by $s^{BS}_k$. Assumed that the required computing resources of all users are configured as a constant $f^{UE}$. The total number of actual computing resources for the MEC server is denoted by $S$. The system model of MEC networks is illustrated in Fig. 1.

The coverage probability of BSs is expressed as \cite{10Andrews}
\begin{equation}
p_c(T,\alpha)=\mathbb{P}[SINR>T]
\label{eq1},
\tag{1}
\end{equation}
where $T$ is the threshold of the signal-to-interference-and-noise ratio (SINR), $\alpha$ is the path loss exponent.

Based on the results in \cite{11Cao,12Zhang,13Xiang}, the SINR of the $k-\text{th}$ BS is expressed as
\begin{equation}
SINR_k = \frac{hr_k^{-\alpha}P_k}{\sigma^2+I_k}
\label{eq2a},
\tag{2a}
\end{equation}
\begin{equation}
I_k=\sum\limits_{m\in B,m\neq k}P_mR_{mk}^{-\alpha},
\label{eq2b}
\tag{2b}
\end{equation}
where $h$ is the channel fading assumed to be an exponentially distributed random variable with parameter $\mu (\mu>0)$ \cite{14Ge}, $\sigma^2$ is the noise power in wireless channels.

As a consequence, the coverage radius $r_k$ is derived as
\begin{equation}
\begin{gathered}
SIN{{R}_{k}}=\frac{hr_{k}^{-\alpha }{{P}_{k}}}{{{\sigma }^{2}}+{{I}_{k}}}=T\Rightarrow {{r}_{k}}={{(\frac{h{{P}_{k}}}{T({{\sigma }^{2}}+{{I}_{k}})})}^{\frac{1}{\alpha }}}
\end{gathered}
\label{eq3}.
\tag{3}
\end{equation}

Furthermore, the cumulative distribution function (CDF) of the coverage radius $r_k$ is expressed as

\begin{equation}
\begin{gathered}
{{F}_{{r}_{k}}}(r)=\mathbb{P}({{r}_{k}}<r)
=\mathbb{P}(h<\frac{T({{\sigma }^{2}}+{{I}_{k}}){{r}^{\alpha }}}{{{P}_{k}}})\hfill \\
\;\;\;\;\;\;\;\;\;\;\;=1-{{e}^{-\frac{\mu T({{\sigma }^{2}}+{{I}_{k}}){{r}^{\alpha }}}{{{P}_{k}}}}} \hfill
\end{gathered}
\label{eq4}.
\tag{4}
\end{equation}

Moreover, the probability density function (PDF) of the coverage radius $r_k$ is expressed as

\begin{equation}
\begin{gathered}
{{f}_{{r}_{k}}}(r)=\frac{d{{F}_{{r}_{k}}}(r)}{dr}=\frac{\alpha \mu T({{\sigma }^{2}}+{{I}_{k}}){{r}^{\alpha -1}}}{{{P}_{k}}}{{e}^{-\frac{\mu T({{\sigma }^{2}}+{{I}_{k}}){{r}^{\alpha }}}{{{P}_{k}}}}} \hfill \\
\end{gathered}
\label{eq5}.
\tag{5}
\end{equation}

In the end, the required computing resources of the $k-\text{th}$ BS is derived as

\begin{small}
\begin{equation}
\begin{gathered}
{{f}^{BS}_{k}}=\int_{0}^{r_M}{f^{UE}\rho {{f}_{{r}_{k}}}(r)2\pi rdr} \hfill \\
\;\;\;\;\;=2f^{UE}\rho \pi \frac{\alpha \mu T({{\sigma }^{2}}+{{I}_{k}})}{{{P}_{k}}}\int_{0}^{r_M}{{{r}^{\alpha }}}{{e}^{-\frac{\mu T({{\sigma }^{2}}+{{I}_{k}})}{{{P}_{k}}}{{r}^{\alpha }}}}dr \hfill
\end{gathered}
\label{eq6}.
\tag{6}
\end{equation}
\end{small}

\section{Allocation solution of Computing Resource}
Based on the potential game theory, a computing resource allocation solution is proposed in this section to reduce energy consumption and improve computing resource efficiency in MEC networks. The computing resource allocation solution includes two parts: the first part is the power control scheme based on the potential game theory and the second part is the computing resource allocation scheme based on linear programming.

\subsection{Power Control Scheme Based on Potential Game}
\subsubsection{Game Formulation} In this paper the power control problem is modeled as an exact potential game model. In this exact potential game model, related utility function of BSs and potential function of the MEC networks are shown below.

The proposed game model is denoted as $\Gamma =\{B ,\mathsf{P}={{\{{{P}_{k}}\}}_{k\in B }}$, ${{\{{{u}_{k}}\}}_{k\in B }}\}$, where $B$ is the set of players, $\mathsf{P}$ is the strategy vector of which the element $P_k$ denotes the transmit power of the player $k$. For the MEC networks, the power control problem can be described as
\begin{equation}
{{\mathsf{P}}^{*}}=\underset{\mathsf{P} }{\mathop{\arg \max }}\,(\Phi (\mathsf{P}))
\label{eq7},
\tag{7}
\end{equation}
where $\Phi (\mathsf{P})$ is the potential function of the MEC networks, which represents the attainable
maximum required computing resources considering interfering BSs.

Potential game is a common game in communication networks. A game can be regarded as the potential game if the influence on the global utility caused by the change in players' strategy is modeled as a single global function. Such single global function is regarded as the potential function.

The expression of an exact potential game in \cite{15Zhu} is given as
\begin{equation}
\Phi ({{t}^{\prime}_{k }},{{t}_{-k}})-\Phi ({{t}_{k}},{{t}_{-k}})=u({{t}^{\prime}_{k }},{{t}_{-k}})-u({{t}_{k}},{{t}_{-k}})
 \label{eq8},
\tag{8}
\end{equation}
where $t_k$ is the strategy of player $k$, $t_{-k}$ is the strategy of all players except $k$. $\Phi ({{t}_{k}},{{t}_{-k}})$ is the potential function of MEC networks, which denotes the overall benefit of the MEC networks. As the player's individual utility function, $u({{t}_{k}},{{t}_{-k}})$ denotes individual benefit of each BS. According to (8), the increment of the potential function of the MEC networks is equal to the increment of the individual utility function of one player, caused by the change in the strategy of the player.

Based on \cite{16Qi}, the individual utility function represent the difference between the benefit and cost of the MEC networks. The benefit of the $k-\text{th}$ BS in the MEC networks is denoted as the attainable maximum required computing resources with no interference, which is expressed as
\begin{equation}
{{f}^{BS}_{k,\max }}=2f^{UE}\rho \pi \frac{\alpha \mu T{{\sigma }^{2}}}{{{P}_{k}}}\int_{0}^{r_M}{{{r}^{\alpha }}}{{e}^{-\frac{\mu T{{\sigma }^{2}}}{{{P}_{k}}}{{r}^{\alpha }}}}dr
\label{eq9}.
\tag{9}
\end{equation}

The cost of the $k-\text{th}$ BS consists of two parts. One is the reduction of the required computing resources caused by introducing the interference from the $m-\text{th}$ BS ($m\ne k$), denoted as $I_{k,m}$. The other part is the reduction of the required computing resources considering the interference received by the $m-\text{th}$ BS from the $k-\text{th}$ BS, denoted as $I_{m,k}$. Based on (6), $I_{k,m}$ is expressed as

\begin{equation}
\begin{gathered}
{{I}_{k,m}}\!=\!2f^{UE}\rho \pi \frac{\alpha \mu T}{{{P}_{k}}}({{\sigma }^{2}}\!\int_{0}^{{{r}_{M}}}\!{{{r}^{\alpha }}}{{e}^{-\frac{\mu T{{\sigma }^{2}}}{{{P}_{k}}}{{r}^{\alpha }}}}\!dr\! \hfill \\
\;\;\;\;\;\;\;-\!({{\sigma }^{2}}+{{P}_{m}}R_{mk}^{-\alpha })\!\int_{0}^{{{r}_{M}}}{{{r}^{\alpha }}}{{e}^{-\frac{\mu T({{\sigma }^{2}}+{{P}_{m}}R_{mk}^{-\alpha })}{{{P}_{k}}}{{r}^{\alpha }}}}\!dr)\hfill
\end{gathered}
\label{eq10},
\tag{10}
\end{equation}
where $P_m$ is the transmit power of the $m-\text{th}$ BS, and $R_{mk}$ is the distance between the $m-\text{th}$ BS with the $k-\text{th}$ BS. And the estimated  reduction of the required computing resources of the $m-\text{th}$ BS caused by the $k-\text{th}$ BS is expressed as
\begin{equation}
\begin{gathered}
{{I}_{m,k}}\!=\!2f^{UE}\rho \pi\frac{\alpha\mu T}{{{P}_{m}}}({{\sigma }^{2}}\!\int_{0}^{r_M}\!{{{r}^{\alpha }}}{{e}^{-\frac{\mu T{{\sigma }^{2}}}{{{P}_{m}}}{{r}^{\alpha }}}}\!dr\! \hfill \\
\;\;\;\;\;\;\;-\!({{\sigma }^{2}}+{{P}_{k}}R_{km}^{-\alpha })\!\int_{0}^{r_M}\!{{{r}^{\alpha }}}{{e}^{-\frac{\mu T({{\sigma }^{2}}+{{P}_{k}}R_{km}^{-\alpha })}{{{P}_{m}}}{{r}^{\alpha }}}}\!dr) \hfill
\end{gathered}
\label{eq11},
\tag{11}
\end{equation}
where $R_{km}$ is the distance between the $k-\text{th}$ BS and the $m-\text{th}$ BS. And $R_{km}=R_{mk}$.

Furthermore, the utility function of the $k-\text{th}$ BS can be expressed as
\begin{equation}
\begin{gathered}
u({{t}_{k}},{{t}_{-k}})\!=\!2f^{UE}\rho \pi \frac{\alpha \mu T{{\sigma }^{2}}}{{{P}_{k}}}\!\int_{0}^{r_M}\!{{{r}^{\alpha }}}{{e}^{-\frac{\mu T{{\sigma }^{2}}}{{{P}_{k}}}{{r}^{\alpha }}}}\!dr\! \hfill \\
\;\;\;\;\;\;\;\;\;\;\;\;\;\;\;-\!\frac{\varepsilon}{K-1}\!\sum\limits_{\begin{smallmatrix}
 m\in B,m\ne k
\end{smallmatrix}}\!{({{I}_{k,m}}+{{I}_{m,k}})}\hfill
\end{gathered}
\label{eq12},
\tag{12}
\end{equation}
where $\varepsilon$ is a constant for balancing required computing resources with interference.

The potential function (i.e. overall benefit of the MEC networks) the weighted sum of individual utility functions of all BSs, expressed as
\begin{small}
\begin{equation}
\begin{gathered}
\Phi (\mathsf{P})=\sum\limits_{k\in B}(
 2f^{UE}\rho\pi \frac{\alpha \mu T{{\sigma }^{2}}}{{{P}_{k}}}\int_{0}^{r_M}\!{{{r}^{\alpha }}}{{e}^{-\frac{\mu T{{\sigma }^{2}}}{{{P}_{k}}}{{r}^{\alpha }}}}dr \hfill \\
\;\;\;\;\;\;\;\;\;-\frac{\varepsilon}{K-1}(b\sum\limits_{m\in B,m\ne k}{{{I}_{k,m}}}\!+\!
(1-b)\sum\limits_{m\in B,m\ne k}{{{I}_{m,k}}})) \hfill
\label{eq13}
\tag{13},
\end{gathered}
\end{equation}
\end{small}
where $b$ is a constant for building exact potential game.

Based on (12) and (13), the game is proved to be an exact potential game in the appendix. Monderer and Shapley demonstrated the theorem that each finite potential game has at least one pure strategy NE \cite{17Mon}. The theorem guarantee the existence of NE for the exact potential games \cite{18Ling}.

\subsubsection{PSO Based Potential Game}
The particle swarm optimization (PSO) algorithm is proposed to solve the potential game in Algorithm 1.

Each particle in PSO algorithm represents a solution to a specific problem. In other words, a particle is a point in a multi-dimensional search space in which we are attempting to find an optimal location with respect to a fitness function \cite{19Ozcan}. Parameters of PSO algorithm  include group number $N$, maximum iteration number $Ger$, inertia weight $\omega$, self-learning factor $c1$, group learning factor $c2$ and search dimension $d$ which is equal to the total number of BSs. The notations in the PSO algorithm are shown in Table I \cite{19Ozcan}.

\begin{table}[h]
        \setlength{\abovecaptionskip}{0.cm}
        \setlength{\belowcaptionskip}{-0.cm}
        \caption{Notation}\label{tab2}
        \centering
        \begin{tabular}{|c|l|}
        \hline
        \textbf{Symbols} &  \textbf{Meanings}   \\
        \hline
        $x$   & A set of positions (states) of N particles \\
        \hline
        $x_i$  & The position of particle $i$, which is a set of BSs's transmit power \\
        \hline
        $U_i$ &	The current fitness of particle $i$ \\
        \hline
        $ppm$ &	The historical optimal position of particle $i$ \\
        \hline
        $Uppm$ & The historical optimal fitness of particle $i$ \\
        \hline
        $gpm$ & The historical optimal position of group \\
        \hline
        $Ugpm$	& The historical optimal fitness of group  \\
        \hline
        \end{tabular}
        \label{table1.}
 \end{table}

\subsection{Computing Resource Allocation Scheme Based on Linear Programming}

In this section, we discuss the classification of the results obtained in the proposed PSO algorithm. The actually allocated computing resources at the $k-\text{th}$ BS is denoted as $s_k$. The computing resources required by the $k-\text{th}$ BS $f_k$ can be calculated based on (6) and the optimal power control scheme. $S$ indicates the total number of computing resources owned by the MEC server. The classification is discussed as follows.

\begin{enumerate}
\item  $\sum\limits_{B }{{{f}^{BS}_{k}}}\le S$, indicates that the total number of computing resources required by all BSs is less than or equal to the total number of actual computing resources.

    Then ${{s}^{BS}_{k}}={{f}^{BS}_{k}},\forall k\in B $, the required computing resources of each BS can be satisfied.
\item  $\sum\limits_{B }{{{f}^{BS}_{k}}}>S$, indicates that the total number of computing resources required by all BSs is more than the total number of actual computing resources.

    To solve this problem, the resource allocation coefficient is proposed in this paper, which denotes the ratio of the actually allocated computing resources to the required computing resources. The resource allocation coefficient is denoted as $Sat=\frac{1}{K}\sum\limits_{k\in B}{\frac{{{s}^{BS}_{k}}}{{{f}^{BS}_{k}}}}$.
    In order to maximize the average computing resource allocation coefficient, a specific linear programming problem is formulated as

\begin{small}
\begin{equation}
\begin{gathered}
 \max\text{ }Sat=\frac{1}{K}\sum\limits_{k\in B }{\frac{{{s}^{BS}_{k}}}{{{f}^{BS}_{k}}}}\text{ }s.t.\left\{
 \begin{array}{ll}
 \sum\limits_{k\in B }{{{s}^{BS}_{k}}}\le S \\
 0\le {{s}^{BS}_{k}}\le {{f}^{BS}_{k}},\forall k\in B  \\
\end{array}\right. \\
\end{gathered}
\label{eq14}.
\tag{14}
\end{equation}
\end{small}

\end{enumerate}

\begin{algorithm}[h]
\setlength{\textfloatsep}{0.1cm}
 \setlength{\floatsep}{0.1cm}
    \caption{PSO algorithm based on potential game}
    \begin{algorithmic}[1]
    \STATE\textbf{Input:} Total number of base stations $K$; maximum transmit power of BSs $P_{max}$; users density $\rho$; unit user required computing resources $f^{UE}$; maximum coverage radius of BSs $r_M$; total number of computing resources $S$.
        \STATE Initialize the swarm randomly
        \FOR{each particle $i$ in the search space}
          \STATE 1) Initialize feasible position and velocity
          \STATE 2) Set position information $ppm$, $Uppm$, $gpm$ and $Ugpm$
          \STATE 3) Set the lower bound and upper bound of each parameter
        \ENDFOR
        \WHILE{maximum iterations is not attained}
          \FOR{each particle $i$}
            \STATE 1) Calculate fitness value $U_i$
            \STATE 2) Update $Uppm$, $ppm$, $Ugpm$ and $gpm$
             \IF{$U_i>Uppm$}
                        \STATE $Uppm=U_i$
                        \STATE $ppm=x_i$
                        \IF{$U_i>Ugpm$}
                          \STATE $Ugpm=U_i$
                          \STATE $gpm=x_i$
                        \ENDIF
                      \ENDIF
          \ENDFOR
          \FOR{each particle $i$}
            \STATE 1) Calculate particle velocity: $v = \omega  * v + c1 * rand() * (ppm - x) + c2 * rand() * (gpm - x)$
            \STATE 2) Update particle position: $x=x+v$
          \ENDFOR
        \ENDWHILE
        \STATE\textbf{Output:} $gpm$ and $Ugpm$
    \end{algorithmic}
\end{algorithm}

\section{Simulations And Result Analysis}

In this section, the proposed computing resource allocation solution is simulated in a grid-based system. The simulation results of the proposed solution are analyzed and compared with the reference solution. Reference solution 1 is the classical equal allocation solution \cite{20Rhee}, reference solution 2 is ${{s}^{BS}_{k}}=\min ({{f}^{BS}_{k}},S/K)$.

The $100m\times100m$ square zone as the simulation scenario. And the main simulation parameters are shown in Table II \cite{21Humar}\cite{22Ge}\cite{23Ge}\cite{24Ge}.

\begin{table}
\setlength{\textfloatsep}{0.1cm}
 \setlength{\floatsep}{0.1cm}
        \caption{Simulation Parameters}\label{tab2}
        \centering
        \begin{tabular}{|c|c|c|c|}
        \hline
        \textbf{Parameter} &  \textbf{Value} &
        \textbf{Parameter} &  \textbf{Value}  \\
        \hline
        $f^{UE}$  & 1 CPU cycles/bit & $\varepsilon$	& 0.5 \\
        \hline
        $\rho$  & $10^{-2} users/m^2$ &
        $b$	& 0.5 \\
        \hline
        $\mu$ &	1 &$N$ &	6 \\
        \hline
        $T$ &	10dB & $Ger$  & 5 \\
        \hline
        $\sigma^2$ & $10^{-15}W$ & $\omega$ & 0.8 \\
        \hline
        $P_{max}$ & 5W & $c1$ & 0.9 \\
        \hline
        $r_M$	& 100m & $c2$ & 0.9  \\
        \hline
        $S$	& 1CPU cycles/bit  & - & -\\
        \hline
        \end{tabular}
        \label{table2.}
 \end{table}

\begin{figure}[h]
\centerline{\includegraphics[width=0.35\textwidth]{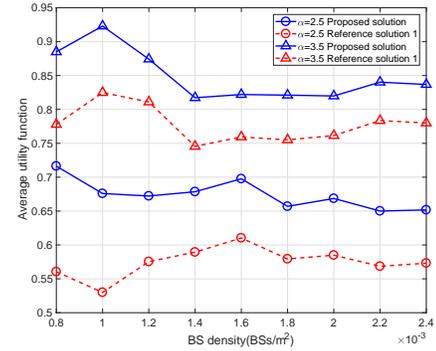}}
\caption{Average utility function with respect to the BS density considering different path loss exponent.}
\label{fig2}
\end{figure}
\begin{figure}[h]
\centerline{\includegraphics[width=0.35\textwidth]{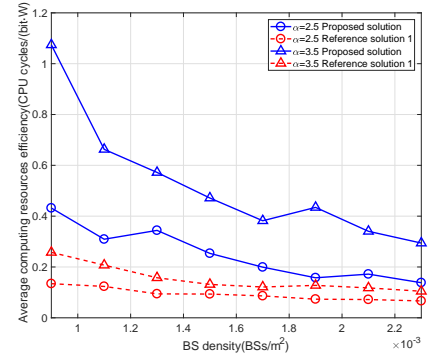}}
\caption{Average computing resource efficiency with respect to the BS density considering different path loss exponent.}
\label{fig3}
\end{figure}
\begin{figure}[h]
\centerline{\includegraphics[width=0.35\textwidth]{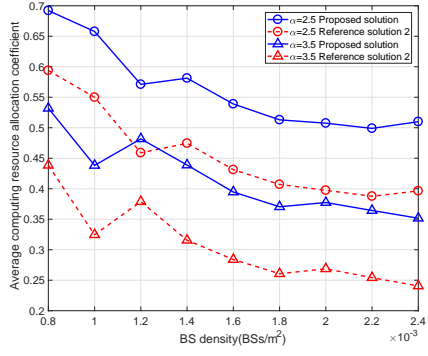}}
\caption{Average computing resource allocation coefficient with respect to the BS density considering different path loss exponent.}
\label{fig4}
\end{figure}

Fig. 2 shows the average utility function with respect to the BS density considering different path loss exponents. The average utility function of the proposed solution is compared with that of the reference solution 1. When the BS density is fixed, the average utility function increases with the increase of the path loss exponent. When the path loss exponent is fixed, the average utility function decreases with the increase of the BS density. The average utility function of the proposed solution is larger than that of the reference solution 1.

Fig. 3 shows the average computing resource efficiency (average required computing resources/transmit power of BSs) with respect to the BS density considering different path loss exponents. The average computing resource efficiency of the proposed solution is compared with that of the reference solution 1. When the BS density is fixed, the average computing resource efficiency increases with the increase of the path loss exponent. When the path loss exponent is fixed, the average computing resource efficiency decreases with the increase of the BS density. The average computing resource efficiency of the proposed solution is larger than that of the reference solution 1.

Fig. 4 shows the computing resource allocation coefficient with respect to the BS density considering different path loss exponents. The computing resource allocation coefficient decreases with the increase of the BS density.

The performance of the proposed solution simulated in Fig. 2, Fig. 3 and Fig. 4 is better than the performance of the reference solutions for the MEC networks.

\section{Conclusion}

A computing resource allocation solution for MEC networks is proposed in this paper. The optimization model is constructed, and the potential game theory is explored to guarantee the convergence of utility function. This solution included the power control scheme based on potential game theory and the resource allocation scheme based on linear programming. Finally, the proposed computing resource allocation solution is evaluated by using a grid-based system. Simulation results show that compare with traditional solutions, the computing resource utilization and energy efficiency of the proposed computing resource allocation solution are significantly improved. So the proposed computing resource allocation solution is applicable under energy saving scene. We hope that the computing resource allocation solution proposed in this paper can promote the development of saving energy and computing resource allocation in MEC networks in the future.

\section*{ACKNOWLEDGMENT}
The authors would like to acknowledge the support from National Key R$\&$D Program of China (2016YFE0133000): EU-China study on IoT and 5G (EXICITING-723227)

\section*{Appendix: Demonstration of Validity for the Proposed Potential Game Model}
Now let us prove that the game is the exact potential game, which is equivalent to prove the equation in (8) holds. Firstly, the proposed potential function in (13) will be decomposed:
\begin{small}
\[\begin{gathered}[h]
\Phi ({{t}_{k}},{{t}_{-k}}) \hfill \\
\!=\!2f^{UE}\rho \pi \alpha \mu T\sum\limits_{B}(
{\frac{{{\sigma }^{2}}}{{{P}_{k}}}\int_{0}^{r_M}{{{r}^{\alpha }}}{{e}^{-\frac{\mu T{{\sigma }^{2}}}{{P}_{k}}{{r}^{\alpha }}}}dr} \hfill \\
\!-\!\varepsilon \frac{b}{K-1}\sum\limits_{m\in B,m\ne k}(\frac{{{\sigma }^{2}}}{{{P}_{k}}}\int_{0}^{r_M}{{{r}^{\alpha }}}{{e}^{-\frac{\mu T{{\sigma }^{2}}}{{{P}_{k}}}{{r}^{\alpha }}}}dr \hfill \\
\!-\!\frac{{{\sigma }^{2}}+{{P}_{m}}R_{mk}^{-\alpha }}{{{P}_{k}}}\!\int_{0}^{r_M}\!{{{r}^{\alpha }}}{{e}^{-\frac{\mu T({{\sigma }^{2}}+{{P}_{m}}R_{mk}^{-\alpha })}{{{P}_{k}}}{{r}^{\alpha }}}}\!dr) \hfill \\
\!-\!\varepsilon \frac{1-b}{K-1}\!\sum\limits_{m\ne k}\!(\frac{{{\sigma }^{2}}}{{{P}_{m}}}\!\int_{0}^{r_M}\!{{{r}^{\alpha }}}{{e}^{-\frac{\mu T{{\sigma }^{2}}}{{{P}_{m}}}{{r}^{\alpha }}}}\!dr \hfill \\
\!-\!\frac{{{\sigma }^{2}}+{{P}_{k}}R_{km}^{-\alpha }}{{{P}_{m}}}\int_{0}^{r_M}{{{r}^{\alpha }}}{{e}^{-\frac{\mu T({{\sigma
}^{2}}+{{P}_{k}}R_{km}^{-\alpha })}{{{P}_{m}}}{{r}^{\alpha }}}}dr)) \hfill \\
\!=\!2f^{UE}\rho \pi \alpha \mu T(\frac{{{\sigma }^{2}}}{{{P}_{k}}}\!\int_{0}^{r_M}{{{r}^{\alpha }}}{{e}^{-\frac{\mu T{{\sigma }^{2}}}{{{P}_{k}}}{{r}^{\alpha }}}}\!dr \hfill \\
\!+\!\varepsilon \frac{b}{K-1}\!\sum\limits_{m\in B,m\ne k}\!{\frac{{{\sigma }^{2}}+{{P}_{m}}R_{mk}^{-\alpha }}{{{P}_{k}}}\!\int_{0}^{r_M}\!{{{r}^{\alpha }}}{{e}^{-\frac{\mu T({{\sigma }^{2}}+{{P}_{m}}R_{mk}^{-\alpha })}{{{P}_{k}}}{{r}^{\alpha }}}}\!dr} \hfill \\
\!+\!\varepsilon \frac{1-b}{K-1}\sum\limits_{m\in B,m\ne k}{\frac{{{\sigma }^{2}}+{{P}_{k}}R_{km}^{-\alpha }}{{{P}_{k}}}\int_{0}^{r_M}{{{r}^{\alpha }}}{{e}^{-\frac{\mu T({{\sigma }^{2}}+{{P}_{k}}R_{km}^{-\alpha })}{{{P}_{m}}}{{r}^{\alpha }}}}dr} \hfill \\
\!-\!\varepsilon \frac{b}{K-1}\sum\limits_{m\ne k}{\frac{{{\sigma }^{2}}}{{{P}_{k}}}\int_{0}^{r_M}{{{r}^{\alpha }}}{{e}^{-\frac{\mu T{{\sigma }^{2}}}{{{P}_{k}}}{{r}^{\alpha }}}}dr} \hfill \\
\!-\!\varepsilon \frac{1-b}{K-1}\sum\limits_{m\ne k}{\frac{{{\sigma }^{2}}}{{{P}_{m}}}\int_{0}^{r_M}{{{r}^{\alpha }}}{{e}^{-\frac{\mu T{{\sigma }^{2}}}{{{P}_{m}}}{{r}^{\alpha }}}}dr}) \hfill \\
\!+\!2f^{UE}\rho \pi \alpha \mu T\sum\limits_{n\in B,n\ne k}(
\frac{{{\sigma }^{2}}+{{P}_{k}}R_{kn}^{-\alpha }}{{{P}_{n}}}
\int_{0}^{r_M}{{{r}^{\alpha }}}{{e}^{-\frac{\mu T({{\sigma }^{2}}+{{P}_{k}}R_{kn}^{-\alpha })}{{{P}_{n}}}{{r}^{\alpha }}}}dr \hfill \\
+\!\varepsilon \frac{1-b}{K-1}\frac{{{\sigma }^{2}}+{{P}_{n}}R_{nk}^{-\alpha }}{{{P}_{k}}}\int_{0}^{r_M}{{{r}^{\alpha }}}{{e}^{-\frac{\mu T({{\sigma }^{2}}+{{P}_{n}}R_{nk}^{-\alpha })}{{{P}_{k}}}{{r}^{\alpha }}}}dr \hfill \\
\!-\!\varepsilon \frac{b}{K-1}\frac{{{\sigma }^{2}}}{{{P}_{n}}}\int_{0}^{r_M}{{{r}^{\alpha }}}{{e}^{-\frac{\mu T{{\sigma }^{2}}}{{{P}_{n}}}{{r}^{\alpha }}}}dr
\!-\!\varepsilon \frac{1-b}{K-1}\frac{{{\sigma }^{2}}}{{{P}_{m}}}\int_{0}^{r_M}{{{r}^{\alpha }}}{{e}^{-\frac{\mu T{{\sigma }^{2}}}{{{P}_{m}}}{{r}^{\alpha }}}}dr
)
\hfill \\
\!+\!2f^{UE}\rho \pi \alpha \mu T\sum\limits_{n\in B,n\ne k}(
\frac{{{\sigma }^{2}}}{{{P}_{n}}}\int_{0}^{r_M}{{{r}^{\alpha }}}{{e}^{-\frac{\mu T{{\sigma }^{2}}}{{{P}_{n}}}{{r}^{\alpha }}}}dr \hfill \\
\!-\!\varepsilon \frac{b}{K-1}\!\sum\limits_{m\ne n,m\ne k}\!(\frac{{{\sigma }^{2}}}{{{P}_{n}}}\int_{0}^{r_M}\!{{{r}^{\alpha }}}{{e}^{-\frac{\mu T{{\sigma }^{2}}}{{{P}_{n}}}{{r}^{\alpha }}}}\!dr \hfill \\
\!-\!\frac{{{\sigma }^{2}}+{{P}_{m}}R_{mn}^{-\alpha }}{{{P}_{n}}}\!\int_{0}^{r_M}\!{{{r}^{\alpha }}}{{e}^{-\frac{\mu T({{\sigma }^{2}}+{{P}_{m}}R_{mn}^{-\alpha })}{{{P}_{n}}}{{r}^{\alpha }}}}\!dr) \hfill \\
\!-\!\varepsilon \frac{1-b}{K-1}\!\sum\limits_{m\ne n,m\ne k}\!(\frac{{{\sigma }^{2}}}{{{P}_{m}}}\!\int_{0}^{r_M}\!{{{r}^{\alpha }}}{{e}^{-\frac{\mu T{{\sigma }^{2}}}{{{P}_{m}}}{{r}^{\alpha}}}}\!dr
\hfill \\
\!-\!\frac{{{\sigma }^{2}}+{{P}_{n}}R_{nm}^{-\alpha }}{{{P}_{m}}}\!\int_{0}^{r_M}\!{{{r}^{\alpha }}}{{e}^{-\frac{\mu T({{\sigma }^{2}}+{{P}_{n}}R_{nm}^{-\alpha })}{{{P}_{m}}}{{r}^{\alpha}}}}\!dr)
) \hfill \\
\end{gathered}\]
\end{small}
\begin{small}
\[\begin{gathered}
\!=\!2f^{UE}\rho \pi \alpha \mu T(\frac{{{\sigma }^{2}}}{{{P}_{k}}}\!\int_{0}^{r_M}{{{r}^{\alpha }}}{{e}^{-\frac{\mu T{{\sigma }^{2}}}{{{P}_{k}}}{{r}^{\alpha }}}}\!dr \hfill \\
\!+\!\varepsilon \frac{1}{K-1}\!\sum\limits_{m\in B,m\ne k}{\frac{{{\sigma }^{2}}+{{P}_{m}}R_{mk}^{-\alpha }}{{{P}_{k}}}\!\int_{0}^{r_M}{{{r}^{\alpha }}}{{e}^{-\frac{\mu T({{\sigma }^{2}}+{{P}_{m}}R_{mk}^{-\alpha })}{{{P}_{k}}}{{r}^{\alpha }}}}\!dr} \hfill \\
\!+\!\varepsilon \frac{1}{K-1}\!\sum\limits_{m\in B,m\ne k}\!{\frac{{{\sigma }^{2}}+{{P}_{k}}R_{km}^{-\alpha }}{{{P}_{k}}}\!\int_{0}^{r_M}\!{{{r}^{\alpha }}}{{e}^{-\frac{\mu T({{\sigma }^{2}}+{{P}_{k}}R_{km}^{-\alpha })}{{{P}_{m}}}{{r}^{\alpha }}}}\!dr} \hfill \\
\!-\!\varepsilon \frac{1}{K-1}\!\sum\limits_{m\in B,m\ne k}\!{\frac{{{\sigma }^{2}}}{{{P}_{k}}}\!\int_{0}^{r_M}\!{{{r}^{\alpha }}}{{e}^{-\frac{\mu T{{\sigma }^{2}}}{{{P}_{k}}}{{r}^{\alpha }}}}\!dr}\hfill \\
\!-\!\varepsilon \frac{1}{K-1}\sum\limits_{m\in B,m\ne k}{\frac{{{\sigma }^{2}}}{{{P}_{m}}}\int_{0}^{r_M}{{{r}^{\alpha }}}{{e}^{-\frac{\mu T{{\sigma }^{2}}}{{{P}_{m}}}{{r}^{\alpha }}}}dr}) \hfill \\
\!+\!2f^{UE}\rho \pi \alpha \mu T\sum\limits_{n\ne k}(
\frac{{{\sigma }^{2}}}{{{P}_{n}}}\int_{0}^{r_M}{{{r}^{\alpha}}}{{e}^{-\frac{\mu T{{\sigma }^{2}}}{{{P}_{n}}}{{r}^{\alpha }}}}dr \hfill \\
\!-\!\varepsilon \frac{b}{K-1}\!\sum\limits_{m\ne n,m\ne k}\!(\frac{{{\sigma }^{2}}}{{{P}_{n}}}\!\int_{0}^{r_M}\!{{{r}^{\alpha }}}{{e}^{-\frac{\mu T{{\sigma }^{2}}}{{{P}_{n}}}{{r}^{\alpha }}}}\!dr\! \hfill \\
-\!\frac{{{\sigma }^{2}}+{{P}_{m}}R_{mn}^{-\alpha }}{{{P}_{n}}}\!\int_{0}^{r_M}\!{{{r}^{\alpha }}}{{e}^{-\frac{\mu T({{\sigma }^{2}}+{{P}_{m}}R_{mn}^{-\alpha })}{{{P}_{n}}}{{r}^{\alpha }}}}\!dr) \hfill \\
\!-\!\varepsilon \frac{1-b}{K-1}\!\sum\limits_{m\ne n,m\ne k}\!(\frac{{{\sigma }^{2}}}{{{P}_{m}}}\!\int_{0}^{r_M}\!{{{r}^{\alpha }}}{{e}^{-\frac{\mu T{{\sigma }^{2}}}{{{P}_{m}}}{{r}^{\alpha }}}}\!dr \hfill \\
\!-\!\frac{{{\sigma }^{2}}+{{P}_{n}}R_{nm}^{-\alpha }}{{{P}_{m}}}\!\int_{0}^{r_M}{{{r}^{\alpha }}}{{e}^{-\frac{\mu T({{\sigma }^{2}}+{{P}_{n}}R_{nm}^{-\alpha })}{{{P}_{m}}}{{r}^{\alpha }}}}\!dr)
) \hfill \\
\end{gathered}\]
\end{small}
Since the items other than the first four items are not related to $k$, we can ignore them in the next calculation.
\begin{small}
\[\begin{gathered}
\Phi ({{t}^{\prime}_{k }},{{t}_{-k}})-\Phi ({{t}_{k}},{{t}_{-k}})
\hfill \\
\!=\!2f^{UE}\rho \pi \alpha \mu T(\frac{{{\sigma }^{2}}}{{{P}^{\prime}_{k}}}\int_{0}^{r_M}{{{r}^{\alpha }}}{{e}^{-\frac{\mu T{{\sigma }^{2}}}{{{P}^{\prime}_{k}}}{{r}^{\alpha }}}}dr \hfill \\
\!+\!\varepsilon \frac{1}{K-1}\sum\limits_{m\in B,m\ne k}{\frac{{{\sigma }^{2}}+{{P}_{m}}R_{mk}^{-\alpha }}{{{P}^{\prime}_{k}}}\int_{0}^{r_M}{{{r}^{\alpha }}}{{e}^{-\frac{\mu T({{\sigma }^{2}}+{{P}_{m}}R_{mk}^{-\alpha })}{{{P}^{\prime}_{k}}}{{r}^{\alpha }}}}dr\;\;\;\;\;\;\;\;\;\;\;\;\;\;\;\;\;\;\;\;\;\;\;\;\;\;\;\;\;\;\;\;\;\;\;\;\;\;\;} \hfill \\
\!+\!\varepsilon \frac{1}{K-1}\sum\limits_{m\in B,m\ne k}{\frac{{{\sigma }^{2}}+{{P}^{\prime}_{k}}R_{km}^{-\alpha }}{{{P}^{\prime}_{k}}}\int_{0}^{r_M}{{{r}^{\alpha }}}{{e}^{-\frac{\mu T({{\sigma }^{2}}+{{P}^{\prime}_{k}}R_{km}^{-\alpha })}{{{P}_{m}}}{{r}^{\alpha }}}}dr                           } \hfill \\
\!-\!\varepsilon \frac{1}{K-1}\sum\limits_{m\in B,m\ne k}{\frac{{{\sigma }^{2}}}{{{P}^{\prime}_{k}}}\int_{0}^{r_M}{{{r}^{\alpha }}}{{e}^{-\frac{\mu T{{\sigma }^{2}}}{{{P}^{\prime}_{k}}}{{r}^{\alpha }}}}dr} \hfill \\
\!-\!\varepsilon \frac{1}{K-1}\!\sum\limits_{m\ne k}\!{\frac{{{\sigma }^{2}}}{{{P}_{m}}}\!\int_{0}^{r_M}\!{{{r}^{\alpha }}}{{e}^{-\frac{\mu T{{\sigma }^{2}}}{{{P}_{m}}}{{r}^{\alpha }}}}\!dr}) \hfill \\
\!-\!(2f^{UE}\rho \pi \alpha \mu T(\frac{{{\sigma }^{2}}}{{{P}_{k}}}\int_{0}^{r_M}{{{r}^{\alpha }}}{{e}^{-\frac{\mu T{{\sigma }^{2}}}{{{P}_{k}}}{{r}^{\alpha }}}}dr \hfill \\
\!+\!\varepsilon \frac{1}{K-1}\sum\limits_{m\in B,m\ne k}{\frac{{{\sigma }^{2}}+{{P}_{m}}R_{mk}^{-\alpha }}{{{P}_{k}}}\int_{0}^{r_M}{{{r}^{\alpha }}}{{e}^{-\frac{\mu T({{\sigma }^{2}}+{{P}_{m}}R_{mk}^{-\alpha })}{{{P}_{k}}}{{r}^{\alpha }}}}dr} \hfill \\
\!+\!\varepsilon \frac{1}{K-1}\sum\limits_{m\in B,m\ne k}{\frac{{{\sigma }^{2}}+{{P}_{k}}R_{km}^{-\alpha }}{{{P}_{k}}}\int_{0}^{r_M}{{{r}^{\alpha }}}{{e}^{-\frac{\mu T({{\sigma }^{2}}+{{P}_{k}}R_{km}^{-\alpha })}{{{P}_{m}}}{{r}^{\alpha }}}}dr}\hfill \\
\!-\!\varepsilon \frac{1}{K-1}\!\sum\limits_{m\ne k}\!{\frac{{{\sigma }^{2}}}{{{P}_{k}}}\!\int_{0}^{r_M}\!{{{r}^{\alpha }}}{{e}^{-\frac{\mu T{{\sigma }^{2}}}{{{P}_{k}}}{{r}^{\alpha }}}}\!dr} \hfill \\
\!-\!\varepsilon \frac{1}{K-1}\!\sum\limits_{m\ne k}{\frac{{{\sigma }^{2}}}{{{P}_{m}}}\!\int_{0}^{r_M}\!{{{r}^{\alpha }}}{{e}^{-\frac{\mu T{{\sigma }^{2}}}{{{P}_{m}}}{{r}^{\alpha }}}}\!dr})) \hfill \\
\!=\!u({{t}^{\prime}_{k }},{{t}_{-k}})-u({{t}_{k}},{{t}_{-k}}) \hfill \\
\end{gathered}\]
\end{small}

\end{document}